\documentclass[journal]{IEEEtran}

\usepackage[dvips]{color}
\usepackage{graphicx}
\usepackage{amsmath}

\begin{document}

\title{Intercept Behavior Analysis of Industrial Wireless Sensor Networks in the Presence of Eavesdropping Attack}

\markboth{IEEE Transactions on Industrial Informatics (accepted to appear)}{Yulong Zou \MakeLowercase{\textit{et al.}}: Intercept Behavior Analysis of Industrial Wireless Sensor Networks in the Presence of Eavesdropping Attack}

\author{Yulong Zou,~\IEEEmembership{Senior Member,~IEEE}, and Gongpu Wang

\thanks{Manuscript received July 14, 2014; revised December 14, 2014; accepted January 30, 2015. Paper no. TII-14-0732.}

\thanks{Copyright (c) 2009 IEEE. Personal use of this material is permitted. However, permission to use this material for any other purposes must be obtained from the IEEE by sending a request to pubs-permissions@ieee.org.}

\thanks{This work was partially supported by the Priority Academic Program Development of Jiangsu Higher Education Institutions, the National Natural Science Foundation of China (Grant Nos. 61302104 and 61401223), the Scientific Research Foundation of Nanjing University of Posts and Telecommunications (Grant No. NY214001), and the Natural Science Foundation of Jiangsu Province (Grant No. BK20140887).}

\thanks{Y. Zou (corresponding author) is with the School of Telecommunications and Information Engineering, Nanjing University of Posts and Telecommunications, Nanjing, Jiangsu 210003, China (e-mail: yulong.zou@njupt.edu.cn).}

\thanks{G. Wang (corresponding author) is with the School of Computer and Information Technology, Beijing Jiaotong University, Beijing, China (email: gpwang@bjtu.edu.cn).}

}

\maketitle

\begin{abstract}
This paper studies the intercept behavior of an industrial wireless sensor network (WSN) consisting of a sink node and multiple sensors in the presence of an eavesdropping attacker, where the sensors transmit their sensed information to the sink node through wireless links. Due to the broadcast nature of radio wave propagation, the wireless transmission from the sensors to the sink can be readily overheard by the eavesdropper for interception purposes. In an information-theoretic sense, the secrecy capacity of the wireless transmission is the difference between the channel capacity of the main link (from sensor to sink) and that of the wiretap link (from sensor to eavesdropper). If the secrecy capacity becomes non-positive due to the wireless fading effect, the sensor's data transmission could be successfully intercepted by the eavesdropper and an intercept event occurs in this case. However, in industrial environments, the presence of machinery obstacles, metallic frictions and engine vibrations makes the wireless fading fluctuate drastically, resulting in the degradation of the secrecy capacity. As a consequence, an optimal sensor scheduling scheme is proposed in this paper to protect the legitimate wireless transmission against the eavesdropping attack, where a sensor with the highest secrecy capacity is scheduled to transmit its sensed information to the sink. Closed-form expressions of the probability of occurrence of an intercept event (called intercept probability) are derived for the conventional round-robin scheduling and the proposed optimal scheduling schemes. Also, an asymptotic intercept probability analysis is conducted to provide an insight into the impact of the sensor scheduling on the wireless security. Numerical results demonstrate that the proposed sensor scheduling scheme outperforms the conventional round-robin scheduling in terms of the intercept probability.

\end{abstract}


\IEEEpeerreviewmaketitle

\section{Introduction}

\IEEEPARstart {W}{ireless} sensor networks (WSNs) were initially motivated by the military for battlefield surveillance [1], and now are further developed for various industrial applications such as the assembly line monitoring and manufacturing automation for the sake of improving the factory efficiency, reliability, and productivity [2], [3], which are referred to as the industrial WSNs [4]-[6]. In industrial applications, the real-time communications among the spatially distributed sensors should satisfy strict security and reliability requirements [7]. The failure of ensuring the security and reliability of the sensed information transmissions may cause an outage of the production line, a damage of the factory machine, or even the loss of workers' lives. Moreover, in industrial environments, the machinery obstacles, metallic frictions, engine vibrations and equipment noise are hostile to the radio propagation and certainly adversely affect the performance of wireless transmissions.

In industrial WSNs, due to the broadcast nature of radio propagation, the wireless medium is open to be accessed by both authorized and unauthorized users, leading WSNs to be more vulnerable to the eavesdropping attack than wired sensor networks, where communicating nodes are physically connected with wire cables and a node without being connected is unable to access for illegal activities. To be specific, as long as an eavesdropper hides in the industrial WSNs, the legitimate wireless transmissions among the sensors can be readily overheard by the eavesdropper, which may decode its tapped transmissions and violate the confidentiality of the sensors' information communications [8]. Therefore, it is of importance to investigate the protection of industrial WSNs against the eavesdropping attack.

Traditionally, the cryptographic techniques were exploited to protect the wireless communications against eavesdropping, which typically rely on secret keys and can prevent an eavesdropper with limited computational capability from intercepting the data transmission between wireless sensors. However, an eavesdropper with unlimited computing power is still able to crack the encrypted data communications with the aid of exhaustive key search (known as the brute-force attack) [9], [10]. Moreover, the secret key distribution and agreement between the wireless sensors exhibit numerous vulnerabilities and further increase the security risk. To this end, physical-layer security is emerging as a promising paradigm for secure communications by exploiting the physical characteristics of wireless channels, which can effectively protect the confidentiality of communication against the eavesdropping attack, even with unlimited computational power [11].

The physical-layer security work was pioneered by Shannon in [12] and extended by Wyner in [13], where an information-theoretic framework was established by developing achievable secrecy rates for a classical wiretap channel model consisting of one source, one destination and an eavesdropper. In [14], the so-called \emph{secrecy capacity} was shown as the difference between the channel capacity of the main link from source to destination and that of the wiretap link from source to eavesdropper. If the secrecy capacity becomes non-positive (i.e., the channel capacity of the main link becomes less than that of the wiretap link), the eavesdropper will succeed in intercepting the source message and an intercept event is considered to occur in this case [15], [16]. This implies that increasing the secrecy capacity can effectively decrease the probability that the eavesdropper successfully intercepts the source message. However, the secrecy capacity of wireless transmission is severely limited due to the wireless fading effect. Moreover, the presence of machinery obstacles, metallic frictions and engine vibrations in industrial environments makes the wireless fading fluctuate drastically, resulting in a further degradation of the secrecy capacity.

To overcome this limitation, considerable research efforts have been devoted to improving the secrecy capacity of the wireless transmission through the artificial noise generation [17]-[19]. The artificial noise aided security approaches allow the legitimate transmitters to generate a specifically designed interfering signal (called artificial noise) such that only the eavesdropper is adversely affected by the artificial noise, while the intended receiver remains unaffected. This leads to a degradation of the wiretap link in terms of the channel capacity without affecting the channel capacity of the main link, resulting in an increased secrecy capacity. In [17] and [18], the authors considered the use of multiple antennas for generating the artificial noise and showed that the number of antennas at the legitimate transmitter should be more than that at the legitimate receiver for the sake of ensuring that the main link is unaffected by the artificial noise. Additionally, in [19], Goeckel investigated the employment of cooperative relays for the artificial noise generation and demonstrated a significant security improvement in terms of the secrecy capacity.

It is pointed out that although the artificial noise approaches in [17]-[19] can effectively enhance the wireless secrecy capacity, additional power resources are consumed for generating the artificial noise to confuse the eavesdropper. To this end, Zou \emph{et al.} proposed the multiuser scheduling scheme in [20] for improving the wireless physical-layer security without any additional power cost and showed the security improvement of cognitive radio networks in terms of the secrecy capacity and intercept probability. In this paper, we examine the sensor scheduling in an industrial WSN consisting of a sink node and multiple sensors in the presence of an eavesdropper, differing from the multiuser scheduling for cognitive radio networks as studied [20]. More specifically, in industrial WSNs, the wireless channel is complicated due to the machinery obstacles, metallic frictions and engine vibrations. This motivates us to consider the use of a complex fading model (i.e., Nakagami model) for characterizing the industrial wireless channel, instead of a simpler Rayleigh fading model used in [20].

{{The sensor scheduling to be studied in this paper exhibits some advantages over the conventional relay selection [15], [16] and the artificial noise methods [17]-[19] in terms of reducing the system implementation complexity and saving the power resource. Specifically, in [15] and [16], additional network nodes were introduced and employed for relaying the transmissions between the source and destination, which are refereed to as relay nodes. Given multiple relay nodes available, the authors of [15] and [16] investigated the relay selection for wireless security enhancement, where the relay node that can achieve the highest secrecy against eavesdropping is chosen as the ``best" relay to assist the source-destination transmissions. Although the relay selection studied in [15] and [16] improves the wireless physical-layer security, it relies on additional relay nodes and requires complex synchronization among spatially distributed relays, resulting in extra system complexity. In addition, the artificial noise methods were devised in [17]-[19] to improve wireless security by generating a sophisticatedly-designed artificial noise for confusing the eavesdropper only without affecting the legitimate destination. This, however, costs additional energy resources for the artificial noise generation, compared to the sensor scheduling, where a sensor with the highest secrecy against eavesdropping is scheduled for data transmission without consuming any additional energy resources. Since wireless sensors are usually powered with limited batteries, the energy becomes one of the most precious resources in industrial WSNs, which makes the sensor scheduling more attractive than the conventional artificial noise methods from the energy saving perspective.}}

The main contributions of this paper are summarized as follows. First, an optimal sensor scheduling scheme is proposed for protecting the industrial wireless transmission {{against}} the eavesdropping attack, where a sensor with the highest secrecy capacity is selected to transmit its sensed information to the sink. The conventional round-robin scheduling is also considered as a benchmark. Second, closed-form expressions of the intercept probability for the conventional round-robin scheduling and the proposed optimal sensor scheduling schemes are derived in Nakagami fading environments. Third, an asymptotic intercept probability analysis is conducted and the diversity order of the proposed scheduling scheme is shown as the sum of Nakagami shaping factors of the main links from the sensors to the sink. Finally, numerical results show the advantage of the proposed sensor scheduling scheme over the conventional round-robin scheduling in terms of the intercept probability.

The remainder of this paper is organized as follows. Section II presents the system model of an industrial WSN in the presence of eavesdropping attack and describes the conventional round-robin scheduling as well as the proposed optimal sensor scheduling schemes. Next, in Section III, the closed-form intercept probability expressions of the conventional round-robin scheduling and proposed optimal scheduling schemes are derived in Nakagami fading environments. In Section IV, we present the asymptotic intercept probability analysis, followed by Section V, where the numerical intercept probability comparison between the conventional and proposed sensor scheduling schemes are presented. Finally, Section VI provides some concluding remarks.

\section{Sensor Scheduling in Industrial WSNs}

\subsection{System Model}
\begin{figure}
  \centering
  {\includegraphics[scale=0.6]{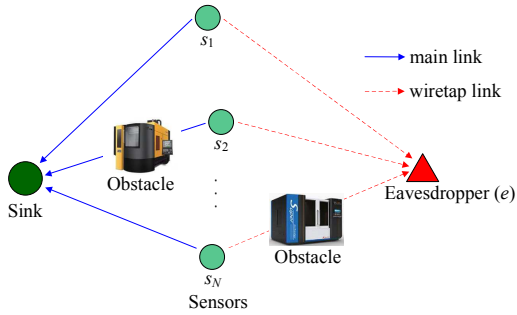}\\
  \caption{An industrial WSN consisting of a sink and $N$ sensors in the presence of an eavesdropper (\emph{e}).}\label{Fig1}}
\end{figure}
As shown in Fig. 1, we consider an industrial WSN consisting of a sink node and $N$ sensors in the presence of an eavesdropper, where all nodes are assumed with single antenna and the solid and dash lines represent the main link and wiretap link, respectively. Notice that the eavesdropper of Fig. 1 could be either an illegitimate user or a legitimate user who is interested in tapping other users' data information. For notational convenience, $N$ sensors are denoted by ${\cal{S}}=\{s_i|i=1,2,\cdots,N\}$. As illustrated in Fig. 1, the presence of machinery obstacles, metallic frictions and engine vibrations in industrial environments is hostile to the radio propagation, which makes the wireless fading fluctuate drastically. We thus consider the use of Nakagami fading model for characterizing both the main channel and wiretap channel. It is pointed out that the Nakagami model is more complex than other fading models (e.g., Rayleigh fading, etc.), which is widely used in literature [21], [22].

In the industrial WSN of Fig. 1, $N$ sensors communicate with the sink using an orthogonal multiple access method such as the time division multiple access (TDMA) and orthogonal frequency division multiple access (OFDMA). When a sensor (e.g., $s_i$) is scheduled to transmit its data to the sink over a channel (e.g., a time slot in TDMA or an OFDM subcarrier in OFDMA), the eavesdropper attempts to intercept the information transmitted from $s_i$. Traditionally, given an orthogonal channel, a node with the highest data throughput is typically selected among $N$ sensors to access the given channel and to communicate with the sink, which aims at maximizing the transmission capacity without considering the eavesdropping attack. By contrast, this paper is focused on improving the wireless physical-layer security with the aid of sensor scheduling. In order to effectively defend against the eavesdropping attack, the sensor scheduling should take into account the channel state information (CSI) of both the main channel and wiretap channel, differing from the traditional scheduling method, where only the CSI of main channel is considered for the throughput maximization. In this paper, we assume that the CSIs of both the main channel and wiretap channel are available, which is an assumption commonly used in the physical-layer security literature (e.g., [14]-[20]).

\subsection{Conventional Round-Robin Scheduling}
For comparison purposes, let us first examine the conventional round-robin scheduling as a benchmark, where $N$ sensors take turns in accessing a given channel and thus each sensor has an equal chance to transmit its sensed data to the sink. Without any loss of generality, we consider that $s_i$ is scheduled to transmit its signal $x_i$ ($E(|{x_i}{|^2}) = 1$) with power $P_i$ and rate $R_i$, where $R_i$ is specified to the maximum achievable rate (also known as the channel capacity) from $s_i$ to the sink, which guarantees that the ergodic capacity is achieved by the legitimate transmission.

{{It needs to be pointed out that the sensed information $x_i$ could be different types of data for different sensors. For example, the $N$ sensors of Fig. 1 may be used to detect and monitor different aspects of an industrial plant environment, including the machine motion, temperature, moisture, pressure, and so on. The sensor data may also be obtained by exploiting the collaboration between multiple sensors for distributed state estimation [23]-[25]. Although the sensors may generate different types of data, their data streams are assumed with the same priority in accessing the wireless channel for transmission in this paper. However, in practice, different data types may have different quality of service (QoS) requirements. For example, some sensors may have time-critical data with a strict real-time requirement, which should be assigned with a higher priority in accessing the wireless channel than the others. It is of interest to consider different QoS requirements for different sensor data, which is out of the scope of this paper and regarded as future work.}}

Thus, we can express the received signal at sink as
\begin{equation}\label{equa1}
{y_s} = \sqrt {P_i} {h_{is}}{x_i} + {n_s},
\end{equation}
where ${h_{is}}$ is a fading coefficient of the main channel from $s_i$ to the sink and $n_s$ represents the zero-mean additive white Gaussian noise (AWGN) with variance $N_0$. Using the Shannon capacity formula [12] and (1), we can obtain the channel capacity of main link from $s_i$ to sink as
\begin{equation}\label{equa2}
{C_s}(i) = {\log _2}\left(1 + \frac{{|{h_{is}}{|^2}P_i}}{{{N_0}}}\right),
\end{equation}
where $i \in {\cal{S}}$. Meanwhile, due to the open nature of wireless medium, the eavesdropper can overhear the signal transmitted from $s_i$ and attempts to decode $x_i$ from its overheard signal. Following the physical-layer security literature [11]-[20], the eavesdropper is assumed to have the perfect knowledge of legitimate transmissions from $s_i$ to sink, including the coding and modulation scheme, encryption algorithm and secret key, except that the source signal $x_i$ is confidential. Hence, the signal overheard at eavesdropper $e$ is given by
\begin{equation}\label{equa3}
{y_{e}} = \sqrt {P_i} {h_{ie}}{x_i} + {n_{e}},
\end{equation}
where ${h_{ie}}$ is a fading coefficient of the wiretap channel from $s_i$ to the eavesdropper and $n_{e}$ represents the zero-mean AWGN with variance $N_0$. Using (3), we can similarly obtain the channel capacity of wiretap link from $s_i$ to eavesdropper $e$ as
\begin{equation}\label{equa4}
{C_{e}}(i) = {\log _2}\left(1 + \frac{{|{h_{ie}}{|^2}P_i}}{{{N_{0}}}}\right).
\end{equation}
As discussed in [13] and [14], the secrecy capacity is shown as the difference between the channel capacity of main link and that of wiretap link. Therefore, in the presence of eavesdropping attack, the secrecy capacity of wireless transmission from $s_i$ to sink can be obtained as
\begin{equation}\label{equa5}
{C_{\textrm{secrecy}}}(i) = {C_s}(i) - {C_{{e}}}(i),
\end{equation}
where ${C_s}(i)$ and ${C_{{e}}}(i)$ are given by (2) and (4), respectively.

\subsection{Proposed Optimal Sensor Scheduling}
This subsection presents an optimal sensor scheduling scheme to maximize the secrecy capacity of the legitimate transmission. Naturally, a sensor with the highest secrecy capacity should be chosen and scheduled to transmit its data to the sink. Hence, from (5), the optimal sensor scheduling criterion is given by
\begin{equation}\label{equa6}
\begin{split}
{\textrm{Optimal User}} &= \arg \mathop {\max }\limits_{i \in {\cal{S}}} {C_{\textrm{secrecy}}}(i) \\
&= \arg \mathop {\max }\limits_{i \in {\cal{S}}} \dfrac{{1 + \frac{{|{h_{is}}{|^2}P_i}}{{{N_0}}}}}{{1 + \frac{{|{h_{ie}}{|^2}P_i}}{{{N_0}}}}},
\end{split}
\end{equation}
where $\cal{S}$ represents the set of $N$ sensors. {{It is observed from (6) that the channel state information (CSI) (i.e., $|h_{is}|^2$ and $|h_{ie}|^2$) of each sensor is required for determining the optimal sensor, which can be obtained by using classic channel estimation methods [26]-[28]. More specifically, each sensor may first estimate its own CSI through channel estimation and then transmits the estimated CSI to the sink. After collecting all the sensors' CSI, the sink can readily determine the optimal sensor and notify the whole network.}} Thus, in the presence of an eavesdropper, the secrecy capacity of legitimate transmissions relying on the proposed sensor scheduling scheme can be obtained from (6) as
\begin{equation}\label{equa7}
{C^{\textrm{proposed}}_{\textrm{secrecy}}}=\mathop {\max }\limits_{i \in {\cal{S}}} {\log _2}\left( {\dfrac{{1 + \frac{{|{h_{is}}{|^2}P_i}}{{{N_0}}}}}{{1 + \frac{{|{h_{ie}}{|^2}P_i}}{{{N_0}}}}}} \right).
\end{equation}
It is pointed out that the wireless link between any two nodes of Fig. 1 is modeled as the Nakagami fading channel. Thus, $|h_{is}|$ and $|h_{ie}|$ are Nakagami distributed random variables with respective shape factors $m_i$ and $k_i$, which, in turn, leads to the fact that $|h_{is}|^2$ and $|h_{ie}|^2$ are Gamma distributed, i.e., $|{h_{is}}{|^2} \sim \Gamma ({m_i},\frac{{\sigma _{is}^2}}{{{m_i}}})$ and $|{h_{ie}}{|^2} \sim \Gamma ({k_i},\frac{{\sigma _{ie}^2}}{{{k_i}}})$, where $\sigma _{is}^2$ and $\sigma _{ie}^2$ are expected values of $|{h_{is}}{|^2}$ and $|{h_{ie}}{|^2}$, respectively. Denoting ${X_{is}} = |{h_{is}}{|^2}$ and ${X_{ie}} = |{h_{ie}}{|^2}$, we can obtain the probability density functions (PDFs) of ${X_{is}}$ and ${X_{ie}}$ as
\begin{equation}\label{equa8}
\begin{split}
&{f_{{X_{is}}}}({x_{is}}) = \frac{1}{{\Gamma ({m_i})}}{(\frac{{{m_i}}}{{\sigma _{is}^2}})^{{m_i}}}x_{is}^{{m_i} - 1}\exp ( - \frac{{{m_i}{x_{is}}}}{{\sigma _{is}^2}})\\
&{f_{{X_{ie}}}}({x_{ie}}) = \frac{1}{{\Gamma ({k_i})}}{(\frac{{{k_i}}}{{\sigma _{ie}^2}})^{{k_i}}}x_{ie}^{{k_i} - 1}\exp ( - \frac{{{k_i}{x_{ie}}}}{{\sigma _{ie}^2}}),
\end{split}
\end{equation}
where $\Gamma ( \cdot )$ denotes Gamma function.

\section{Exact Intercept Probability Analysis over Nakagami Fading Channels}
In this section, we analyze the intercept probability of the conventional round-robin scheduling and the proposed optimal sensor scheduling schemes over Nakagami fading channels.

\subsection{Round-Robin Scheduling}
As shown in [15]-[17], when the secrecy capacity becomes non-positive (i.e., the channel capacity of the main link becomes less than that of the wiretap link), the eavesdropper will succeed in decoding and intercepting the source message and an intercept event is considered to occur in this case.
Hence, given that ${s}_i$ is scheduled to transmit to the sink, the intercept probability of ${{s}}_i$-to-sink transmission is obtained from (5) as
\begin{equation}\label{equa9}
P_{{\mathop{\textrm{int}}} }^i  = \Pr \left[ {C_{{\textrm{secrecy}}} (i) < 0} \right] = \Pr \left[ {C_s (i) < C_e (i)} \right].
\end{equation}
Substituting (2) and (4) into (9) yields
\begin{equation}\label{equa10}
P_{{\mathop{\textrm {int}}} }^i  = \Pr \left( {|h_{is} |^2  < |h_{ie} |^2 } \right),
\end{equation}
where $|h_{is} |^2$ and $|h_{ie} |^2$ are Gamma distributed random variables with respective PDFs shown in (8). Noting that $|h_{is} |^2$ and $|h_{ie} |^2$ are independent of each other and denoting ${X_{is}} = |{h_{is}}{|^2}$ and ${X_{ie}} = |{h_{ie}}{|^2}$, we obtain
\begin{equation}\label{equa11}
\begin{split}
P_{{\mathop{\textrm {int}}} }^i&=\Pr \left( {{X_{is}} < {X_{ie}}} \right)\\
&= \iint\limits_{{{x_{is}} < {x_{ie}}}} {{{f_{{X_{ie}}}}({x_{ie}}){f_{{X_{is}}}}({x_{is}})d{x_{ie}}d{x_{is}}}} \\
&= \int_0^\infty  {\Gamma ({m_i},\frac{{{m_i}}}{{\sigma _{is}^2}}{x_{ie}}){f_{{X_{ie}}}}({x_{ie}})d{x_{ie}}} ,
\end{split}
\end{equation}
where ${f_{{X_{ie}}}}({x_{ie}})$ is given by (8) and $\Gamma ({m_i},\frac{{{m_i}}}{{\sigma _{is}^2}}{x_{ie}})$ denotes the lower incomplete gamma function as given by
\begin{equation}\nonumber
\Gamma ({m_i},\frac{{{m_i}}}{{\sigma _{is}^2}}{x_{ie}}) = \int_0^{\frac{{{m_i}}}{{\sigma _{is}^2}}{x_{ie}}} {\frac{1}{{\Gamma ({m_i})}}{x^{{m_i} - 1}}\exp ( - x)dx}.
\end{equation}
One can observe from (11) that the intercept probability of the ${{s}}_i$-to-sink transmission is independent of the transmit power $P_i$ and noise variance $N_0$, which implies that increasing the transmit power cannot improve the wireless physical-layer security. This also motivates us to explore new approaches (e.g., sensor scheduling) for improving the wireless security against the eavesdropping attack. In the round-robin scheduling scheme, $N$ sensors take turns in transmitting to the sink and thus the intercept probability of the round-robin scheduling is the mean of $N$ sensors' intercept probabilities, yielding
\begin{equation}\label{equa12}
P_{\textrm{int}}^{\textrm{round}} = \frac{1}{N}\sum\limits_{i = 1}^N {P_{{\mathop{\textrm {int}}} }^i},
\end{equation}
where $N$ is the number of sensors and $P_{{\mathop{\textrm {int}}} }^i$ is given by (11). It is noted that numerical intercept probability results of the round-robin scheduling scheme can be readily determined by using (12).

\subsection{Optimal Sensor Scheduling}
This subsection derives an exact intercept probability of the proposed optimal sensor scheduling scheme over Nakagami fading channels. As aforementioned, an intercept event occurs when the secrecy capacity becomes non-positive. Hence, using (7), we obtain the intercept probability of proposed sensor scheduling scheme as
\begin{equation}\label{equa13}
\begin{split}
{P^{\textrm{proposed}}_{{\mathop{\textrm {int}}} }} &= \Pr \left( {{C^{\textrm{proposed}}_{\textrm{secrecy}}} < 0} \right) \\
&= \Pr \left[ {\mathop {\max }\limits_{i \in {\cal{S}}} {{\log }_2}\left( {\frac{{1 + \frac{{|{h_{is}}{|^2}P_i}}{{{N_0}}}}}{{1 + \frac{{|{h_{ie}}{|^2}P_i}}{{{N_0}}}}}} \right) < 0} \right]\\
&=\Pr \left[ {\mathop {\max }\limits_{i \in {\cal S}} \left( {\frac{{N_0  + |h_{is} |^2 P_i }}{{N_0  + |h_{ie} |^2 P_i }}} \right) < 1} \right],
\end{split}
\end{equation}
where $\cal S$ denotes the set of $N$ sensors. Noting that for different sensors $i \in {\cal S}$, random variables $|h_{is} |^2$ and $|h_{ie}|^2$ are independent of each other, we can simplify (13) as given by
\begin{equation}\label{equa14}
\begin{split}
{P^{\textrm{proposed}}_{{\mathop{\textrm {int}}} }}
&= \prod\limits_{i = 1}^N {\Pr \left( {\frac{{N_0  + |h_{is} |^2 P_i }}{{N_0  + |h_{ie} |^2 P_i }} < 1} \right)}\\
&= \prod\limits_{i = 1}^N {\Pr \left( {|{h_{is}}{|^2} < |{h_{ie}}{|^2}} \right)}\\
&=\prod\limits_{i = 1}^N {P_{{\mathop{\textrm {int}}} }^i},
\end{split}
\end{equation}
where $P_{{\mathop{\textrm {int}}} }^i$ is given by (11). So far, we have derived the closed-form intercept probability expressions of both the conventional round-robin scheduling and the proposed optimal scheduling schemes in Nakagami fading environments. It is worth mentioning that although the intercept probability expressions given by (12) and (14) can be used to conduct numerical performance evaluation, it fails to provide an insight into the impact of the number of sensors on the intercept probability. Therefore, the following section presents an asymptotic intercept probability analysis to characterize the diversity order performance.

\section{Asymptotic Intercept Probability Analysis of Sensor Scheduling}
In this section, we analyze the asymptotic intercept probability of both the conventional round-robin and proposed optimal scheduling schemes.
\subsection{Conventional Round-Robin Scheduling}
Let us first consider the conventional round-robin scheduling as baseline. For notational convenience, the average channel gains of the main link and wiretap link (i.e., $\sigma_{is}^2$ and $\sigma_{ie}^2$) are represented by $\sigma_{is}^2=\alpha_{is}\sigma^2_{m}$ and $\sigma_{ie}^2=\alpha_{ie}\sigma^2_{e}$, where $\sigma^2_{m}$ and $\sigma^2_{e}$ are called the reference channel gains of the main link and the wiretap link, respectively. Moreover, let $\lambda_{me}$ represent the ratio of $\sigma^2_{m}$ to $\sigma^2_{e}$, i.e., $\lambda_{me}=\sigma^2_{m}/\sigma^2_{e}$, which is referred to as the main-to-eavesdropper ratio (MER) throughout this paper. We will examine the asymptotic intercept probability as MER $ \lambda_{me} \to \infty$. Following [15], the diversity gain of the conventional round-robin scheduling scheme is expressed as
\begin{equation}\label{equa15}
{d_{{\textrm{round}}}} = -\mathop {\lim }\limits_{{\lambda_{me}} \to \infty } \dfrac{{\log (P^{\textrm{round}}_{{\textrm{int}}})}}{{\log({\lambda _{me}})}},
\end{equation}
where $P^{\textrm{round}}_{{\textrm{int}}}$ is given by (12). We first rewrite $P_{{\mathop{\textrm {int}}} }^i$ from (11) as
\begin{equation}\label{equa16}
\begin{split}
P_{{\mathop{\textrm {int}}} }^i &= \Pr \left( {{X_{is}} < {X_{ie}}} \right) \\
&= \Pr \left( {\frac{{{X_{is}}}}{{\sigma _e^2}} < \frac{{{X_{ie}}}}{{\sigma _e^2}}} \right),
\end{split}
\end{equation}
where $\sigma^2_{e} > 0$ is the reference channel gain of the wiretap link. Denoting ${Y_{is}} = \frac{{{X_{is}}}}{{\sigma _e^2}}$, ${Y_{ie}} = \frac{{{X_{ie}}}}{{\sigma _e^2}}$, ${\alpha _{is}} = \frac{{\sigma _{is}^2}}{{\sigma _m^2}}$ and ${\alpha _{ie}} = \frac{{\sigma _{ie}^2}}{{\sigma _e^2}}$, we readily obtain the PDFs of ${Y_{is}} $ and ${Y_{ie}} $ from (8) as
\begin{equation}\label{equa17}
\begin{split}
&{f_{{Y_{is}}}}({y_{is}}) = \frac{1}{{\Gamma ({m_i})}}{(\frac{{{m_i}}}{{{\alpha _{is}}{\lambda _{me}}}})^{{m_i}}}y_{is}^{{m_i} - 1}\exp ( - \frac{{{m_i}{y_{is}}}}{{{\alpha _{is}}{\lambda _{me}}}})\\
&{f_{{Y_{ie}}}}({y_{ie}}) = \frac{1}{{\Gamma ({k_i})}}{(\frac{{{k_i}}}{{{\alpha _{ie}}}})^{{k_i}}}y_{ie}^{{k_i} - 1}\exp ( - \frac{{{k_i}{y_{ie}}}}{{{\alpha _{ie}}}}).
\end{split}
\end{equation}
Using the PDFs of ${Y_{is}} $ and ${Y_{ie}} $ given by (17) and letting $\lambda_{me} \to \infty$, we obtain (18) at the top of the following page, where $\exp ( - \frac{{{m_i}{y_{is}}}}{{{\alpha _{is}}{\lambda _{me}}}}) = 1$ for $\lambda_{me} \to \infty$ is used.
\begin{figure*}
\begin{equation}\label{equa18}
\begin{split}
P_{{\mathop{\textrm {int}}} }^i&= \int_0^\infty  {\frac{1}{{\Gamma ({k_i})}}{(\frac{{{k_i}}}{{{\alpha _{ie}}}})^{{k_i}}}y_{ie}^{{k_i} - 1}\exp ( - \frac{{{k_i}{y_{ie}}}}{{{\alpha _{ie}}}})d{y_{ie}}\int_0^{{y_{ie}}} {\frac{1}{{\Gamma ({m_i})}}{(\frac{{{m_i}}}{{{\alpha _{is}}{\lambda _{me}}}})^{{m_i}}}y_{is}^{{m_i} - 1}\exp ( - \frac{{{m_i}{y_{is}}}}{{{\alpha _{is}}{\lambda _{me}}}})d{y_{is}}} } \\
&= \int_0^\infty  {\frac{1}{{\Gamma ({k_i})}}{(\frac{{{k_i}}}{{{\alpha _{ie}}}})^{{k_i}}}y_{ie}^{{k_i} - 1}\exp ( - \frac{{{k_i}{y_{ie}}}}{{{\alpha _{ie}}}})d{y_{ie}}\int_0^{{y_{ie}}} {\frac{1}{{\Gamma ({m_i})}}{(\frac{{{m_i}}}{{{\alpha _{is}}{\lambda _{me}}}})^{{m_i}}}y_{is}^{{m_i} - 1}d{y_{is}}} } \\
&= \int_0^\infty  {\frac{1}{{{m_i}\Gamma ({k_i})\Gamma ({m_i})}}{(\frac{{{k_i}}}{{{\alpha _{ie}}}})^{{k_i}}}{(\frac{{{m_i}}}{{{\alpha _{is}}{\lambda _{me}}}})^{{m_i}}}y_{ie}^{{m_i} + {k_i} - 1}\exp ( - \frac{{{k_i}{y_{ie}}}}{{{\alpha _{ie}}}})d{y_{ie}}}.
\end{split}
\end{equation}
\end{figure*}
Using (18) and letting $\lambda_{me} \to \infty$, we can further rewrite $P_{{\mathop{\textrm {int}}} }^i $ as
\begin{equation}\label{equa19}
P_{{\mathop{\textrm {int}}} }^i = \frac{{\zeta ({m_i},{k_i},{\alpha _{ie}})}}{{{m_i}\Gamma ({m_i})}}{\left( {\frac{{{m_i}}}{{{\alpha _{is}}}}} \right)^{{m_i}}} \cdot {\left( {\frac{1}{{{\lambda _{me}}}}} \right)^{{m_i}}},
\end{equation}
where $\zeta ({m_i},{k_i},{\alpha _{ie}})$ is given by
\begin{equation}\nonumber
\begin{split}
\zeta ({m_i},{k_i},{\alpha _{ie}}) = &\int_0^\infty  {\frac{1}{{\Gamma ({k_i})}}{{(\frac{{{k_i}}}{{{\alpha _{ie}}}})}^{{k_i}}}y_{ie}^{{m_i} + {k_i} - 1}}\\
&\quad\quad\times \exp ( - \frac{{{k_i}{y_{ie}}}}{{{\alpha _{ie}}}})d{y_{ie}}.
\end{split}
\end{equation}
Combining (12) and (19) yields
\begin{equation}\label{equa20}
\begin{split}
P_{{\mathop{\textrm{int}}} }^{\textrm{round}}  = &\frac{1}{N}\sum\limits_{i = 1}^N {\frac{{\zeta (m_i ,k_i ,\alpha _{ie} )}}{{m_i \Gamma (m_i )}}\left( {\frac{{m_i }}{{\alpha _{is} }}} \right)^{m_i }}\\
&\quad\quad\quad\times\left( {\frac{1}{{\lambda _{me} }}} \right)^{m_i  - \mathop {\min }\limits_{i \in {\cal S}} m_i }  \\
&\quad \cdot \left( {\frac{1}{{\lambda _{me} }}} \right)^{\mathop {\min }\limits_{i \in {\cal S}} m_i }.\\
\end{split}
\end{equation}
Substituting (20) into (15) gives
\begin{equation}\label{equa21}
d_{{\textrm{round}}}  = \mathop {\min }\limits_{i \in {\cal S}} m_i,
\end{equation}
which shows that the intercept probability of the conventional round-robin scheduling decreases exponentially with $\mathop {\min }\limits_{i \in {\cal S}} m_i$ as ${{\lambda _{me}}} \to \infty$, where $m_i$ is the Nakagami shaping factor of the channel from sensor $s_i$ to sink. In other words, the diversity gain of the round-robin scheduling with $N$ sensors is determined by the sensor with the smallest Nakagami shaping factor. This also means that upon increasing the number of sensors, the wireless security of the conventional round-robin scheduling scheme would not improve, and even degrades.

\subsection{Proposed Optimal Sensor Scheduling}
In this subsection, we analyze the diversity gain of proposed optimal sensor scheduling through an asymptotic intercept probability analysis for $ \lambda_{me} \to \infty$. Similarly to (15), the diversity gain of proposed sensor scheduling scheme is given by
\begin{equation}\label{equa22}
{d_{{\textrm{proposed}}}} = -\mathop {\lim }\limits_{{\lambda_{me}} \to \infty } \dfrac{{\log (P^{\textrm{proposed}}_{{\textrm{int}}})}}{{\log({\lambda _{me}})}},
\end{equation}
where $P^{\textrm{round}}_{{\textrm{int}}}$ is given by (14). Substituting (19) into (14), we have
\begin{equation}\label{equa23}
\begin{split}
{P^{\textrm{proposed}}_{{\mathop{\textrm{int}}} }} =& \prod\limits_{i = 1}^N {\left[ {\frac{{\zeta ({m_i},{k_i},{\alpha _{ie}})}}{{{m_i}\Gamma ({m_i})}}{{\left( {\frac{{{m_i}}}{{{\alpha _{is}}}}} \right)}^{{m_i}}}} \right]}\\
&\cdot {\left( {\frac{1}{{{\lambda _{me}}}}} \right)^{\sum\limits_{i = 1}^N {{m_i}} }},
\end{split}
\end{equation}
for $ \lambda_{me} \to \infty$. Substituting $P^{\textrm{round}}_{{\textrm{int}}}$ from (23) into (22) gives
\begin{equation}\label{equa24}
{d_{{\textrm{proposed}}}} = \sum\limits_{i = 1}^N {{m_i}},
\end{equation}
which shows that for ${{\lambda _{me}}} \to \infty$, the intercept probability of proposed optimal sensor scheduling scheme decreases exponentially with $\sum\nolimits_{i = 1}^N {{m_i}} $, i.e., the diversity order of $\sum\nolimits_{i = 1}^N {{m_i}} $ is achieved by the proposed scheduling scheme. One can also observe from (24) that the diversity order of proposed sensor scheduling scheme is the sum of Nakagami shaping factors $m_i$ of the main links from $N$ sensors to sink. Thus, as the number of sensors $N$ increases, the diversity order of proposed sensor scheduling scheme increases accordingly. In other words, increasing the number of sensors can significantly decrease the intercept probability of the proposed scheduling scheme. By contrast, the number of sensors even has a negative impact on the security performance of the conventional round-robin scheduling, as implied from (21). This further confirms the advantage of the proposed optimal scheduling over the round-robin scheduling.

\section{Numerical Results and Discussions}
In this section, we present numerical results on the intercept probability of the conventional round-robin scheduling and the proposed optimal sensor scheduling schemes. To be specific, the numerical intercept probability results of the two scheduling schemes are obtained through using (12) and (14). The wireless link between any two network nodes in Fig. 1 is modeled by the Nakagami fading channel, where the Nakagami shaping factors of both the main link and wiretap link are specified to $m_i=k_i=1.5$, unless otherwise stated. Additionally, the fading coefficients $|h_{is}|^2$ and $|h_{ie}|^2$ are assumed to be independent and identically distributed (i.i.d.) random variables, leading to $\alpha_{is}=\alpha_{ie}=1$.

\begin{figure}
  \centering
  {\includegraphics[scale=0.52]{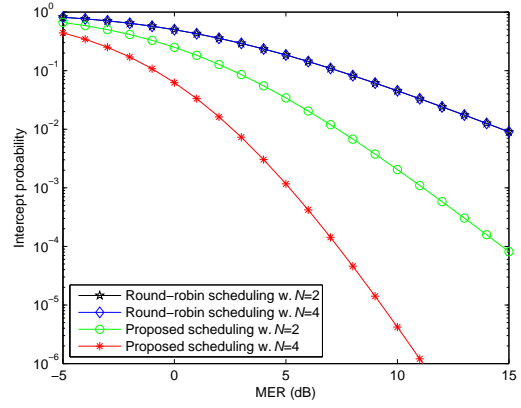}\\
  \caption{Intercept probability versus MER $\lambda_{me}$ of the conventional round-robin scheduling and the proposed optimal scheduling schemes for different number of sensors with $m_i=k_i=1.5$ and $\alpha_{is}=\alpha_{ie}=1$.}\label{Fig2}}
\end{figure}
Fig. 2 shows the intercept probability of the round-robin scheduling and the optimal scheduling schemes by plotting (12) and (14) as a function of MER for different number of sensors $N$. As shown in Fig. 2, as the number of sensors increases from $N=2$ to $N=4$, the intercept probability performance of the conventional round-robin scheduling remains the same, showing that no security benefit is achieved by the round-robin scheduling with an increasing number of sensors. By contrast, upon increasing the number of sensors from $N=2$ to $N=4$, the intercept probability of the proposed optimal scheduling scheme decreases significantly. In addition, Fig. 2 also shows that for both the cases of $N=2$ and $N=4$, the proposed optimal sensor scheduling scheme strictly outperforms the round-robin scheduling in terms of the intercept probability.

\begin{figure}
  \centering
  {\includegraphics[scale=0.52]{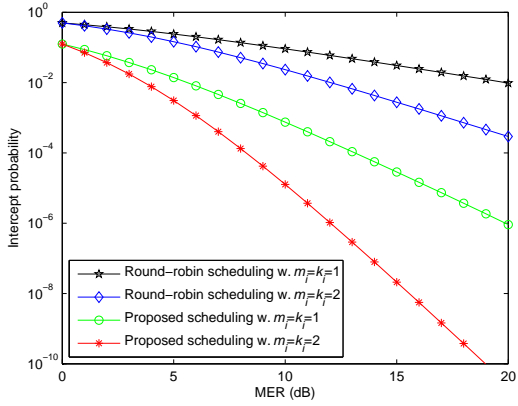}\\
  \caption{Intercept probability versus MER $\lambda_{me}$ of the conventional round-robin scheduling and the proposed optimal scheduling schemes for different Nakagami shaping factors with $N=3$ and $\alpha_{is}=\alpha_{ie}=1$.}\label{Fig3}}
\end{figure}
In Fig. 3, we show the intercept probability versus MER $\lambda_{me}$ of the round-robin scheduling and the optimal scheduling schemes for different Nakagami shaping factors. One can see from Fig. 3 that as the Nakagami shaping factors increase from $m_i=k_i=1$ to $m_i=k_i=2$, the intercept probabilities of both the round-robin scheduling and the optimal scheduling schemes decrease significantly. Fig. 3 also shows that for both $m_i=k_i=1$ and $m_i=k_i=2$, the intercept probability performance of the proposed optimal scheduling is better than that of the round-robin scheduling.

\begin{figure}
  \centering
  {\includegraphics[scale=0.52]{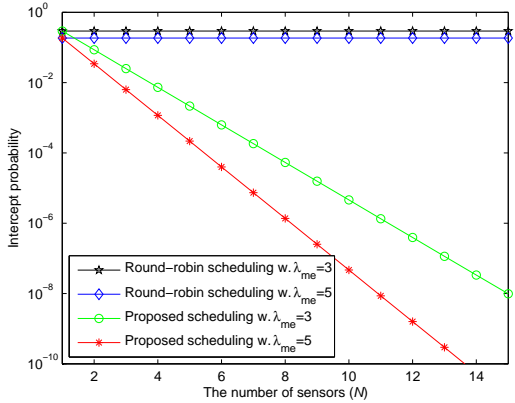}\\
  \caption{Intercept probability versus the number of sensors $N$ of the conventional round-robin scheduling and the proposed optimal scheduling schemes for different MER values with $m_i=k_i=1.5$ and $\alpha_{is}=\alpha_{ie}=1$.}\label{Fig3}}
\end{figure}
Fig. 4 illustrates the intercept probability versus the number of sensors of the conventional round-robin scheduling and the proposed optimal scheduling schemes for different MER values. It can be observed from Fig. 4 that for both the cases of $\lambda_{me}=3$ and $\lambda_{me}=5$, the intercept probability of the conventional round-robin scheduling keeps unchanged, as the number of sensors increases. By contrast, with an increasing number of sensors, the intercept probability performance of the optimal scheduling scheme is significantly improved, showing the security advantage of the proposed scheduling approach over the conventional round-robin scheduling.

\begin{figure}
  \centering
  {\includegraphics[scale=0.52]{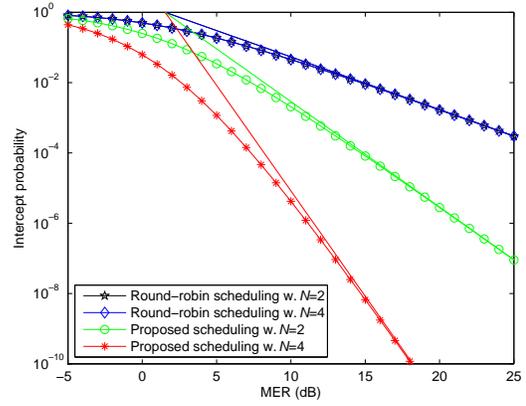}\\
  \caption{Comparison between the exact and asymptotic intercept probabilities of the round-robin scheduling and the proposed optimal scheduling schemes for different number of sensors with $m_i=k_i=1.5$ and $\alpha_{is}=\alpha_{ie}=1$.}\label{Fig3}}
\end{figure}
Fig. 5 shows the comparison between the exact and asymptotic intercept probabilities of the round-robin scheduling and the proposed optimal scheduling schemes for different number of sensors $N$. One can see from Fig. 5 that the intercept probability of the round-robin scheduling corresponding to $N=2$ is the same as that corresponding to $N=4$, further showing that no security benefit is achieved by the round-robin scheduling as the number of sensors increases. It is also observed from Fig. 5 that for both the round-robin scheduling and proposed optimal scheduling schemes, the exact and asymptotic intercept probability results match well with each other in high MER region, implying the tightness of asymptotic intercept probability analysis for $\lambda_{me} \to \infty$. As shown in Fig. 5, the slopes of intercept probability curves of proposed optimal scheduling become much steeper, as the number of sensors $N$ increases from $N=2$ to $N=4$. This means that with an increasing number of sensors, the intercept probability of proposed optimal scheduling scheme decreases at much faster speed as $\lambda_{me} \to \infty$, confirming the advantage of exploiting the optimal sensor scheduling for defending against the eavesdropping attack.

\section{Conclusion and Future Work}
In this paper, we investigated the use of sensor scheduling to improve the physical-layer security of industrial WSNs against the eavesdropping attack and proposed an optimal sensor scheduling scheme, aiming at maximizing the secrecy capacity of wireless transmissions from sensors to the sink. We also considered the conventional round-robin scheduling as a benchmark. We derived exact closed-form expressions of the intercept probability for both the conventional round-robin scheduling and the proposed optimal scheduling schemes in Nakagami fading environments. An asymptotic intercept probability analysis was also presented to characterize the diversity gains of the round-robin scheduling and the optimal sensor scheduling schemes. Numerical results demonstrated that the proposed optimal scheduling scheme performs better than the conventional round-robin scheduling in terms of the intercept probability. In addition, upon increasing the number of sensors, the intercept probability of the proposed optimal sensor scheduling scheme significantly decreases, showing the physical-layer security enhancement of industrial WSNs.

In the present paper, we only examined the single-antenna case, where each network node is equipped with the single antenna. It is of high interest to extend the results of this paper to a general scenario with multiple antennas for each network node. Also, we have not considered the QoS requirement in the sensor scheduling, where all the sensors are assumed with the same priority and scheduled for data transmissions solely based on their channel state information without considering specific QoS requirements for different sensor data. In practice, some sensors may have time-critical data with a strict real-time requirement, which should be assigned with a higher priority than the others in accessing the wireless channel. Hence, it is highly necessary to explore the QoS guaranteed sensor scheduling, attempting to improve the wireless security while guaranteeing each sensor's specific QoS requirement. Additionally, due to the channel estimation errors, it is impossible to obtain the perfect CSI knowledge for the sensor scheduling. It is of thus interest to investigate the impact of CSI estimation errors on the intercept performance of sensor scheduling. We leave these aforementioned challenging but interesting problems for future work.

\begin{IEEEbiography}[{\includegraphics[width=1in,height=1.25in]{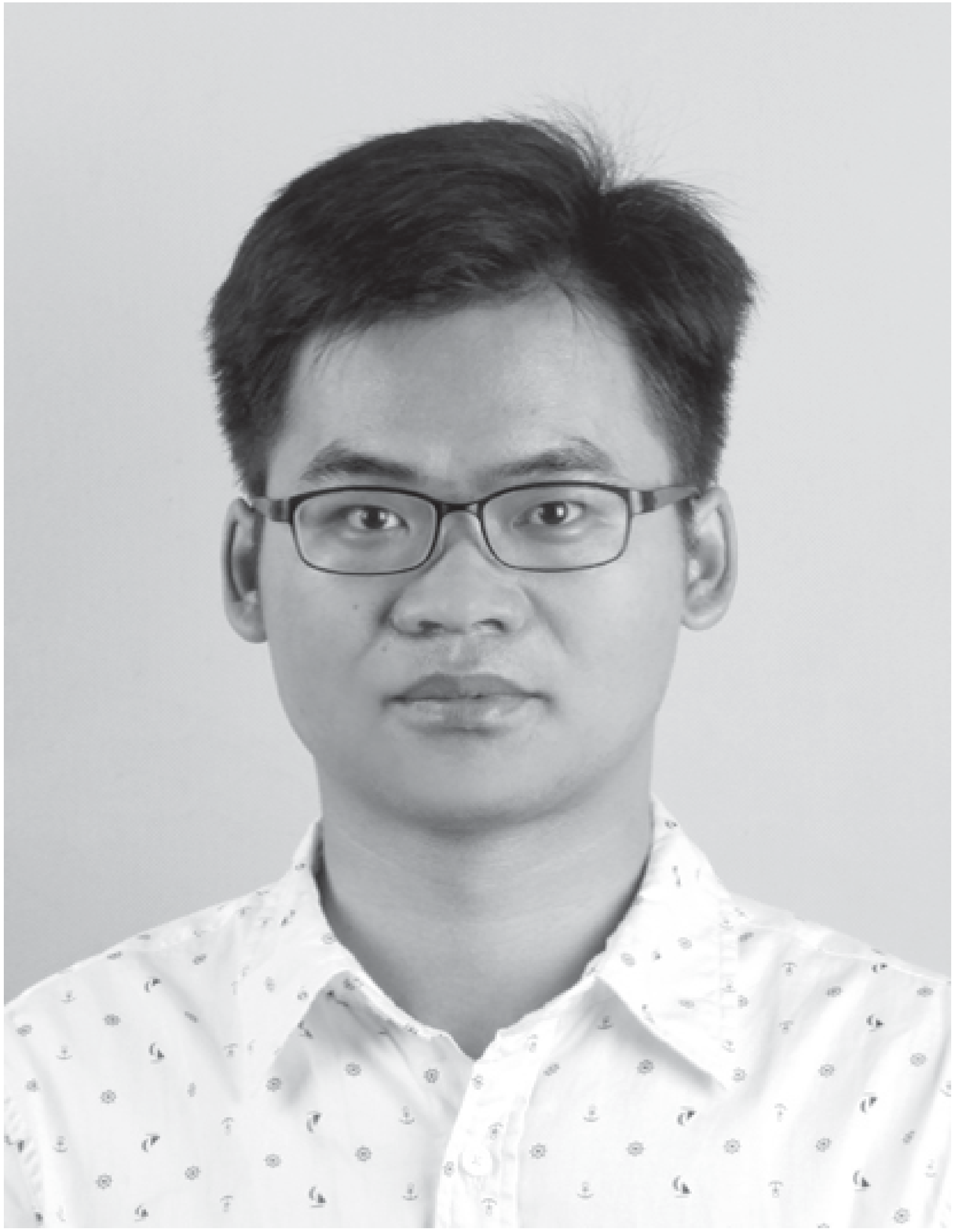}}]{Yulong Zou} (M'12-SM'13) received the B.Eng. degree in Information Engineering from the Nanjing University of Posts and Telecommunications (NUPT), Nanjing, China, in July 2006, the first Ph.D. degree in Electrical Engineering from the Stevens Institute of Technology, New Jersey, the United States, in May 2012, and the second Ph.D. degree in Signal and Information Processing from NUPT, Nanjing, China, in July 2012. He is now a Full Professor and Doctoral Supervisor at the School of Telecommunications and Information Engineering, NUPT, Nanjing, China. His research interests span a wide range of topics in wireless communications and signal processing as well as their industrial applications, including the cooperative communications, cognitive radio, wireless security, and industrial wireless sensor networks.

\end{IEEEbiography}

\begin{IEEEbiography}[{\includegraphics[width=1in,height=1.25in]{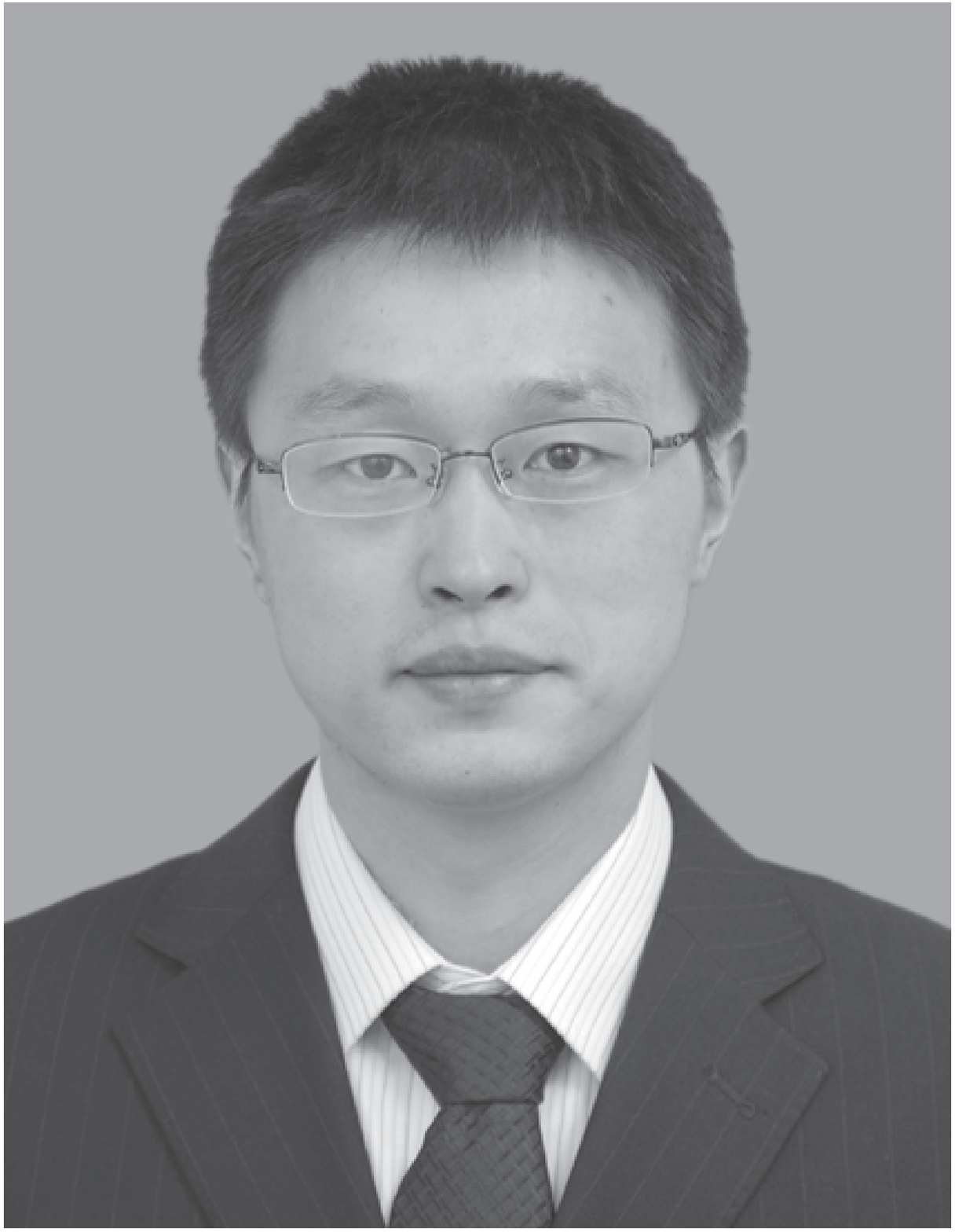}}]{Gongpu Wang} (S'09-M'12) received the B.Eng. degree in Communication Engineering from Anhui University, Hefei, Anhui, China, in 2001, the M.Sc. degree from Beijing University of Posts and Telecommunications, Beijing, China, in 2004. From 2004 to 2007, he was an Assistant Professor in School of Network Education, Beijing University of Posts and Telecommunications. He received Ph.D. degree from University of Alberta, Edmonton, Canada, in 2011. Currently, he is an Associate Professor in School of Computer and Information Technology, Beijing Jiaotong University, China. His research interests include wireless communication theory and signal processing technologies.
\end{IEEEbiography}

\end{document}